# A Survey of Congestion Control Techniques and Data Link Protocols in Satellite Networks[1]

Sonia Fahmy, Raj Jain, Fang Lu[2] and Shivkumar Kalyanaraman
Department of Computer and Information Science
The Ohio State University
Columbus, OH 43210-1277, USA
Phone: (614) 292-3989, Fax: (614) 292-2911
Email: {fahmy, jain, flu, shivkuma}@cis.ohio-state.edu

## Abstract

Satellite communication systems are the means of realizing a global broadband integrated services digital network. Due to the statistical nature of the integrated services traffic, the resulting rate fluctuations and burstiness render congestion control a complicated, yet indispensable function. The long propagation delay of the earth-satellite link further imposes severe demands and constraints on the congestion control schemes, as well as the media access control techniques and retransmission protocols that can be employed in a satellite network. The problems in designing satellite network protocols, as well as some of the solutions proposed to tackle these problems, will be the primary focus of this survey.

# 1 Introduction

Satellite communication systems play a crucial role in the integration of networks of various types and services, and will potentially be used for a wide range of applications. Satellites will continue to play an ever-increasing role in the future of long-range communications [21]. Satellite applications range from navigation, weather monitoring and terrain observations to deep-space exploration.

Satellites of various sizes and capabilities have been launched to serve almost all countries of the world [21]. The main advantages inherent to satellite communications are the broadcasting capability, the full connectivity of stations, the flexibility of station organization, the capacity to support mobile users, and the high transmission quality. In addition, satellite networks can sometimes have the ability to obtain bandwidth on demand.

Several constraints are, however, also inherent in satellite systems. The resources of the satellite communication network, especially the satellite and the earth station, are essentially of a high cost and limited capabilities and flexibility. The most crucial problem, however, is the long propagation delay of the user-satellite links, which renders many traditionally employed schemes such as congestion control schemes, media access control techniques and retransmission protocols inefficient. The integration of various types of traffic on the links further complicates these issues.

---

[1]Submitted to the International Journal of Satellite Communications, 1995.
[2]Fang Lu is currently with Lucent Technologies, New Jersey, USA.

This survey carefully examines some of the solutions proposed to develop efficient protocols for satellite networks, and exposes several problems that are yet to be tackled. The issues pertaining to devising a congestion control mechanism suitable for integrated services traffic on satellite networks are particularly emphasized.

The remainder of this survey is organized as follows. First, the various media access control techniques for satellite networks are examined, exposing several limitations in satellite networks. Then, the use of satellites in interconnecting local area networks is highlighted. The adaptation of retransmission protocols to compensate for the long propagation delay of satellite links is outlined next, followed by a brief discussion of routing schemes. The integration of various types of traffic in Broadband Integrated Services Digital Networks connected via satellites is then explored, and a discussion of the problems in devising a congestion control mechanism for satellite networks concludes the survey.

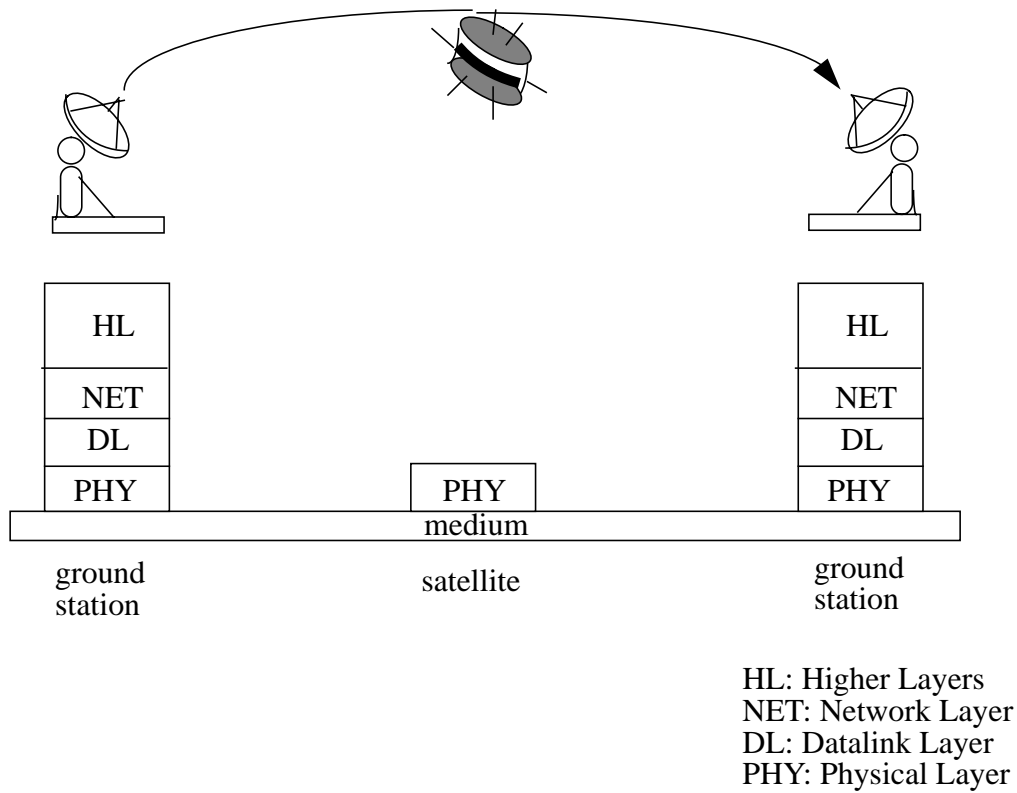

Figure 1: A "Bent-Pipe" Satellite Network

It is noteworthy that both types of satellite communication systems (relay/repeater satellites, as well as processing satellites) are considered. Figures 1 and 2 illustrate two relay satellite (transponder) configurations, while figure 3 shows a processing satellite network (the three figures have been adapted from [14]). Many studies argue that the current trend in satellite systems is towards performing sophisticated processing at the satellite, a technique referred to as on-board processing (OBP). Several schemes developed for OBP systems will be discussed [42, 33, 52, 36, 25, 2, 14, 27].

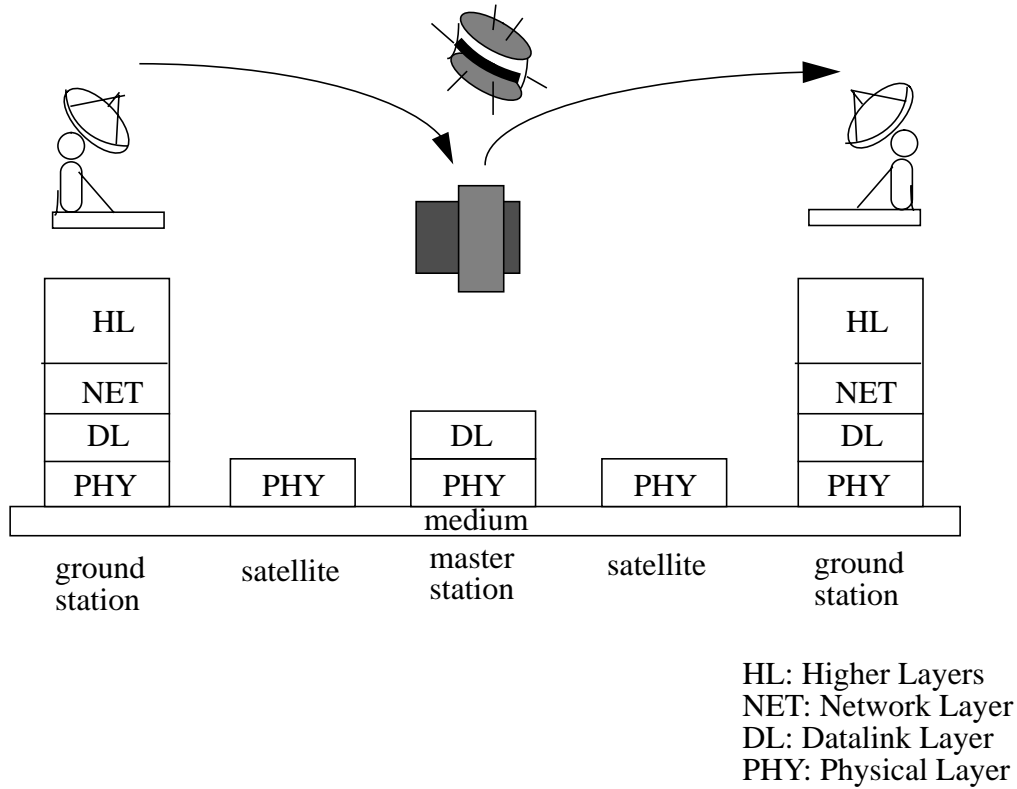

Figure 2: A Two-Hop Satellite Network

## 2 Media Access Control

The development of a multiple access protocol suitable for satellite networks has been the focus of extensive research. A multiple access protocol can be defined as a set of rules for controlling the access to a shared communication channel capacity among various contending users [44].

Earth stations in a satellite network must share the network transmission capacity, and the satellite should be able to handle many simultaneous uplinks and downlinks. Hence some means of controlling access to the transmission medium is needed to provide for an efficient use of its capacity.

Several reservation and random access protocols have been developed for media access control in satellite networks. This section investigates several mechanisms for arbitrating media access, and compares their features and performance. Several problems associated with satellite networks are also briefly examined in this context.

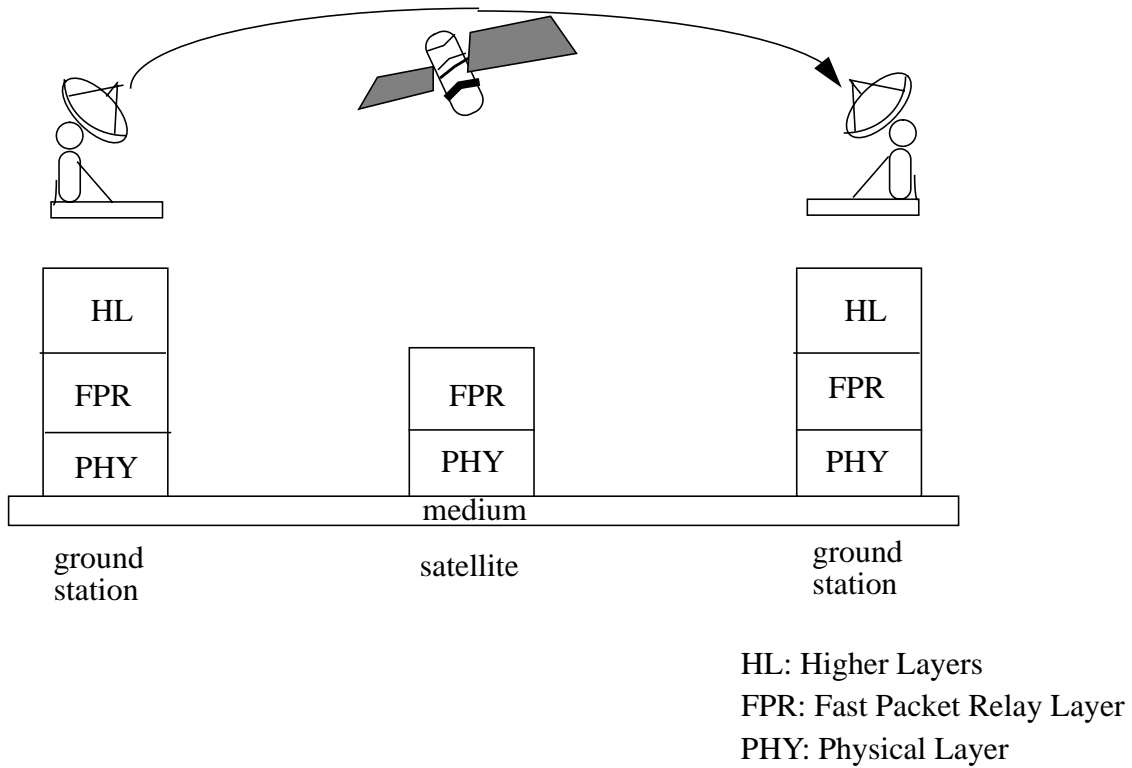

HL: Higher Layers
FPR: Fast Packet Relay Layer
PHY: Physical Layer

Figure 3: A Processing Satellite Network (OBP)

## 2.1 Time Division Multiple Access (TDMA)

Most of the research that investigates multiple access schemes focuses on the Time Division Multiple Access (TDMA) and the slotted-ALOHA schemes. A number of researchers have argued that TDMA is the most suitable mechanism for future broadband integrated services networks, since the efficiency in the use of the satellite channel is high [30, 38]. This section is devoted to examining some of the major issues related to the TDMA access scheme, while the next section investigates slotted-ALOHA.

TDMA allows a station to transmit only during its allotted time interval. Some of the issues pertaining to burst synchronization, time-slot-assignment, fading and data link layer design of TDMA satellite networks are discussed in this section. In addition, several variations on the basic TDMA scheme have been developed and analyzed, and a few of them are presented here.

### 2.1.1 Satellite-switched Time Division Multiple Access

Satellite-switched Time Division Multiple Access (SS-TDMA) is one of the connection schemes of a multibeam satellite communication system, where beam switching is performed in the satellite. This scheme is expected to be used for integrated services networks. Burst syn-chronization and time slot assignment for SS-TDMA are discussed next.

The traditional single-feedback loop for burst synchronization does not achieve good results for many stations in an SS-TDMA multibeam satellite system. In [39], it is proposed that the timing would still be synchronized to the reference station, but the burst synchronization be controlled by the stations for each beam. This still maintains the advantages of the feedback-loop scheme over the closed-loop scheme, while eliminating the performance degradation resulting from simply applying a single feedback loop.

Several sequential and parallel algorithms have been presented for time-slot assignment in SS-TDMA systems with variable-bandwidth beams [11]. The algorithms are based on formulating the time-slot assignment problem as a network-flow problem to find a circulation in a graph model representing the switching system. An SS-TDMA system with variable-bandwidth $M$ uplink beams and $N$ downlink beams is presented as an $M \times N$ traffic matrix where each element $t(i, j)$ denotes the number of time-slots needed to transmit from uplink $i$ to downlink $j$. The problem is mathematically modeled as finding a set of matrices for a given traffic matrix that satisfies some constraints to get maximum traffic handling capacity and minimum overall time slot assignments. The divide-and-conquer approach is used for constructing the algorithms, where a given traffic matrix is decomposed into two smaller traffic matrices. The new algorithms are proved to achieve an improved time complexity [11].

### 2.1.2 TDMA Enhancements and Experiments

Several other modifications to the basic TDMA scheme have been proposed to improve its performance, and make it dynamically adapt to integrated services traffic. A load adaptive TDMA communications link interconnecting broadband multimedia packet streams has been analyzed in [43]. The channels are dynamically assigned to the network stations on the satellite backbone link, and the stations support packetized voice and data message streams. Statistical multiplexing is employed, as well as a variable bit rate packet voice encoding scheme. Algorithms are proposed for allocating the shared backbone channels to the stations, and their performance is analyzed using voice and data packet delays and packet blocking probabilities as metrics. Fairness issues are examined, and effects of propagation delay are discussed. Performance degradations are observed as the allocation delay increases, as well as when the mean "call-on" and "call-off" state durations of data sources decrease.

An on-board switching satellite network employing a modified TDMA technique is described in [42]. The main objective of the scheme is to reduce the end-to-end delay. For up-link access, DTDMA is used where slots are assigned to an earth station on demand, with a minislot for each earth station to transmit its request, so that the access delay is shorter than the fixed TDMA when the traffic is light or unbalanced. For the on-board switching, the multiple input queuing approach is used with separate queues for each down-link destination. This solves the problem where a packet is blocked in the queue even if the out port is available until the packet is at the head of the queue. Thus the switching delay decreases.

Asynchronous TDMA is used in transmission on to the down-links to further reduce the transmission delay. An independent $M/G/1$ queuing system is used as the system model for analysis.

Another adaptive satellite TDMA scheme is the FIFO Ordered Demand Assignment (FODA) access scheme. FODA supports both real-time and non-real-time traffic [10]. The quality of service achieved by a VBR video application (characterized as a Markov process with Bernoulli scene changes) is analyzed and the queuing model of the system is solved in [15]. A conservative approach in bandwidth allocation is adopted, because a low packet loss ratio is desired. Bandwidth allocation is centrally controlled by a station sending a reference burst every 20 milli-seconds. Each FODA frame is divided into a stream sub-frame (allocated at setup time) and a datagram sub-frame. The buffer occupancy and bounds on the maximum access delay are computed using a new algorithm.

The fading of the satellite links under atmospheric attenuation is an important problem, and a method to counteract this fading in TDMA systems is discussed in [26]. When the bandwidth of a satellite link narrows by rain or clouds, the first step to make up this loss is to increase the up-link power up to 8 $dB$ of attenuation. If the fading is still worse than a certain threshold, different transmission bit rates and Forward Error Correction (FEC) coding rates are used. If the total link attenuation exceeds the range of the above countermeasures, the transmission parameters are reset to nominal values. Because of the correlation of the events that cause the link attenuation, many of the links in the system can become attenuated at one time, thus the efficiency of full adaptive resource sharing is constrained. An extra time slot is used in each TDMA frame to compensate for this problem so as to maintain the required system availability [26].

As previously mentioned, the propagation delay is significantly higher in a satellite packet communication network than in a local or wide area network. The isolation of the Media Access Control (MAC) and Logical Link Control (LLC) layers is contrasted to combining them in the satellite packet communication network in [47]. The throughput and the average response time are evaluated for both types, using two MAC protocols: the TDMA-Reservation and the slotted-ALOHA. For TDMA-Reservation (which exhibits a high overhead), combining the two layers shows significantly better performance, while for slotted-ALOHA, no real difference can be observed. A connection-oriented LLC sublayer is employed [47].

In the isolated-type system, LLC frames are handled simply as data frames in the MAC sublayer. Thus in the TDMA-reservation case, not only the transmission of the reservation frame, but also the transmission of various control frames in the LLC sublayer must be executed using the minislots. In contrast, in the combined-type system, the reservation frame must be sent only when the transmission request for the data frame is produced in the LLC sublayer. Various control frames can be sent without reservation. Because of this, the overhead is smaller in the combined type than in the isolated type.

In the case of the slotted-ALOHA communication network, in the combined type, a collision does not occur if the stations transmit the control frames in the same slot. In contrast, in the isolated type, collision always occurs if two or more stations generate control frames in the LLC sublayer in the same slot. By simulation, it can be seen that there is little performance difference except for the case when the average message length is considerably small [47]. The next section discusses the slotted-ALOHA scheme in more depth.

## 2.2 Slotted-ALOHA

In slotted-ALOHA, time on the channel is organized into uniform slots, and transmission is permitted to begin only on a slot boundary. When a station has something to send, it does so, and then listens for some time to detect if it needs to retransmit, due to the occurrence of a collision.

Slotted-ALOHA has been extensively studied as a suitable multiple access scheme for satellite networks. Several studies have attempted to analyze the performance of slotted-ALOHA, and suggest enhancements to achieve better results. A few of them are briefly examined in this section.

### 2.2.1 Capture and Erasure Effects

Slotted-ALOHA can be a high performance multiple access scheme, when one out of several colliding data packets can capture the common receiver (capture effect). If several mobile users can transmit data packets to a central controller (a satellite or a terrestrial base station), the levels of received signals may vary due to fading and shadowing effects. In such cases, the strongest may be able to capture the receiver, so the performance of slotted ALOHA does not decrease as fast as that of the other multiple access schemes when the quality of the transmission channels becomes worse. However, the slotted-ALOHA channel can become unstable if a good retransmission policy is not applied for users with collided packets (backlogged users). In [6], the mean value of the transmission delay after which retransmission occurs is investigated. This value needs to be large enough such that the system does not become unstable due to many retransmissions, but small enough such that the average packet delay is small. The popular retransmission policies can be classified under three broad categories. First, the fixed retransmission probability: this method is simple, but can lead to instability. Second, adaptive strategies: here, retransmission probability is adapted according to the history of the channel. Third, heuristic back-off policies, where the retransmission probability mainly depends on the number of retransmission attempts that have already been performed for the current packet [6].

While the capture effect refers to the situation when one out of several colliding data packets can capture the common receiver due to fading and shadowing effects, the erasure effect refers to the detection of an idle slot when all transmitters are shadowed. Static and dynamic stabilization methods with capture and erasure, are compared both analytically and by simulation in [6]. The finite population model with $N$ users for slotted ALOHA is used. Newly generated packets are treated as previously collided packets (delayed first transmission). The system is described by a homogeneous Markov chain where the number of back-logged users is the state variable. Two algorithms to estimate the actual number of back-logged users are examined: the additive and the multiplicative correcting estimation algorithms. These use information about the outcome of each slot to increase or decrease the estimate. In addition to the analysis, a simulation is performed to compare the various stabilization methods, with an error-prone feedback channel, where not all users get the information about the outcome of a slot. In all cases, dynamic control is established to be much more flexible, and to outperform static policies. The difference between the various dynamic strategies is not dramatic, but a stabilization method based upon the dynamic estimation of the system state is more flexible, and is independent of the total number of users and of the traffic parameter.

### 2.2.2 Slotted-ALOHA Enhancements and Experiments

An enhancement to slotted-ALOHA, namely multi-slotted-ALOHA, has been proposed to improve the throughput in variable length packet signal transmission, by controlling the transmission timing according to the packet-signal length. The transmission timing is set so that the transmission interval is 2 to the power of a non-negative integer $y$, times the shortest transmission interval. Hence, the collision due to partial overlap can be eliminated when the signal length is longer than the slot length, as well as when there is an increase of the simultaneous arrivals and the signal length is shorter than the slot length. The throughput performance of the scheme is formulated for the infinite station model and the finite station model. Simulation results show that the throughput of this new scheme is higher independent of the signal length and the performance is stable against the variation of the signal length distribution. The transmission delay is also analyzed employing the finite station model and equilibrium point analysis, and the multi-slotted-ALOHA also performs better [31].

A further enhancement to the slotted-ALOHA scheme has been proposed in [32]. The users are organized into two groups with different transmitting power, so that the users who have a higher priority are allocated a higher transmitting power and have a higher probability of correct reception when collision occurs. In the analysis of system behavior, a two-dimensional discrete Markovian model is developed, and the results show that the system is not stable. A recursive, distributed two-dimensional auxiliary retransmission control algorithm is presented to stabilize the system. Each back-logged user reschedules the transmissions of the back-logged packets according to the channel feedback, so that the retransmission probabilities at each slot are maintained at an optimal level. Drift analysis is employed for the local model to prove that the parameter choice of this algorithm makes the system stable. The throughput of the group-based random multiple access system is shown to be much higher than the traditional slotted-ALOHA by quantitative analysis.

Another study that examines ALOHA retransmission, classifying the user groups according to transmission power, is presented in [34]. Again, Markov chain model has been developed for the number of back-logged users for each user group classified according to transmission power. The steady state behavior is investigated, and the mean utilization versus access delay is examined. Optimal retransmission strategies are proposed to increase the fairness of the distribution of utilization among the user groups.

Note that communication satellites need to be used to multicast a variety of services, so the access schemes applied to them should accommodate traffic consisting of short, interactive data messages, longer file transfers and voice calls. A large number of geographically dispersed users need to access a variety of multimedia services, and they need to share the communication channel. In [23], the different types of traffic associated with each service are analyzed with several types of multiple-access schemes. Very Small Aperture Terminal (VSAT) satellite networks were employed in the study. VSAT satellite networks are popular because they are small, low-cost, and can be easily installed. The terminals transmit data in packets to the hub station using the multiple access capability of the satellite channel [8]. The multiple-access schemes considered in [23] include a number of variations on the slotted-ALOHA scheme, namely the SREJ-ALOHA and its enhancement, as well as the SREJ-ALOHA/FCFS. Voice/data integration is also investigated by either using a combination of voice channels and data channels, or by taking advantage of the silences in the voice channels to transmit data packets [23].

### 2.2.3 Hybrid Random Access and Reservation Schemes

Several hybrid random access/reservation protocols have been proposed for satellite communications. Traditionally, the only protocols that seemed to combine the benefits of the random access protocols, with those of the reservation protocols, were applied only for single channel multiaccess, which is only effective over a single inbound channel. Future systems are expected to employ on-board processing satellites, with multichannel Frequency Division Multiple Access (FDMA) for uplink access, and single channel Time Division Multiplexing (TDM) for downlink broadcast. A protocol is needed (similar to the demand assignment multiaccess protocols previously considered for circuit switching networks) to be suitable for star-configured packet switched networks with multiple bursty traffic sources, to achieve high throughput and low delays [33].

A hybrid slotted random access protocol has been proposed in [33] for star networks. Many satellite data networks employ star configurations with user terminals communicating with high performance hub earth stations over inbound multiaccess channels and outbound broadcast channels. A new class of multichannel reservation protocols that improve channel utilization through complete sharing of all channels among all users is presented in [33]. Slotted random access is employed, with immediate retransmissions for first-packet and reservation request transmissions, and reservation of consecutive time slots for transmissions of remaining packets in multipacket messages. Through numerical analysis, the method is proved to achieve higher throughput than random access protocols (designed for transmission of individual packets), with lower average delay than other reservation protocols (designed for multiple packets) [33].

The star-configured satellite network considered accomplishes tight inbound synchronization by closed-loop feedback of timing offsets from the hub station. Each user terminal transmits and receives over one inbound and outbound channel at any given time. Inbound and outbound frames are staggered, such that there is sufficient time for the acknowledgment to be received. The multichannel reservation protocol employs random access and immediate retransmission to transmit the first packet of the message to the hub station, with a piggybacked request to reserve capacity for the remaining packets. The hub station maintains a queue of the reservation requests and assigns time slots. Hence user terminals transmit packets using reserved capacity, and retransmission occurs if no acknowledgment is received. Several variations on this basic scheme are possible, depending on the way in which the inbound channels are partitioned. An infinite population Poisson traffic model is used and the throughput capacity and message transmission delays are plotted for various message lengths. The protocol outperforms slotted-ALOHA in most cases [33].

Other researchers have also proposed another hybrid random access/reservation minimum delay protocol for packet satellite communications [53]. Their arrival model is assumed to be a Poisson model and single copy transmission is employed. The delay is minimized by tuning all the system parameters. In particular, the following was observed: a spare reservation should normally, but not always, be made for each packet transmission, all unreserved slots should be filled with a packet rate of one per slot whenever possible, and an optimum balance between transmitting packets and making reservations before transmissions should be maintained.

## 2.3 Code Division Multiple Access (CDMA)

Code Division Multiple Access (CDMA) is a very popular multiple access scheme for the interconnection of cellular networks using Low Earth Orbit (LEO) satellites. In CDMA, a specific coded address waveform is assigned to each carrier. All stations transmitting simultaneously overlap their carrier waveforms on top of each other [21]. CDMA schemes are sometimes called spread spectrum (SS) because a relatively wide bandwidth is used.

Digital addresses are obtained from code generators producing a periodic sequence of symbols. The address sequence of a station is superimposed on the carrier along with the data. If the address is modulated directly on the carrier, the format is referred to as direct-sequence CDMA (DS-CDMA) [21]. The system is referred to as frequency-hopped CDMA (FH-CDMA) if the digital address is used continually to change the frequency of the carrier.

Direct sequence spread spectrum CDMA has a great potential for a high capacity cellular system. The slotted access channel (control channel from the mobile user to the base station) can make use of multipath combining, where if more than one access attempt occurs in a slot, they collide and none of them can be demodulated. The performance of the system with and without multipath combining is analyzed in [20]. The major factors affecting the performance of the channel are collision, blocking and frame erasure rate (FER). These factors can lead to retransmission, which results in a lower throughput and a higher delay. The study attempts to estimate the effective arrival rate in a system without combining using an iterative method. The iteration generates an access rate sequence that converges if the system is stable. Hence, a stability condition is determined in terms of the arrival rate of the new message and the FER. The analysis reveals that the receiving *quality* (FER) is better with the multipath combining, whereas the receiving *ability* is better without multipath combining. For the same low FERs and the same number of demodulators, the access channel effective arrival rates without multipath combining are slightly lower than the ones with multipath combining. Thus, under the same channel utilization, the throughput without the multipath combining can be slightly higher than the one with multipath combining [20].

### 2.3.1 Performance of CDMA in LEO Systems

A multiple low earth orbit (LEO) satellite network is generally known to have several advantages, such as the short propagation delay, low transmission power level and the good coverage [50].

As previously mentioned, CDMA is especially popular in low earth orbit satellite systems used to connect cellular networks. The performance of direct sequence code division multiple access (DS-CDMA) for the multispot beam LEO satellites has been analyzed in [49]. This is worse than the performance of DS-CDMA for cellular terrestrial networks because the channels have different characteristics, such as longer delays from the mobile user to the base station and smaller multipath delay spreads on the satellite channels. The performance of a CDMA system which operates over a LEO channel is analytically derived in that study. Effects such as imperfect power control and dual-order diversity are incorporated to obtain the average probability of error of a single user. In order for DS-CDMA to result in good performance for LEOs, sufficient interleaving must be employed, dual diversity must be used, and a power control system must be implemented so that the standard deviation of the power control error is about 2 $dB$ or less.

The performance of DS-CDMA in a LEO satellite network has also been analyzed in [54]. The system model adopted in this study consists of a constellation of $N$ satellites without inter-satellite links, which are always in sight of all users but can be shadowed. There are both voice and data traffic users which transmit constant length packets and are modeled as two-state discrete-time Markov chains with different parameters. Pure DS spread-spectrum signaling CDMA is used with multiple reception of data packets modeled both as threshold model and graceful degradation model. The numerical results show that for the voice traffic, the blocking probability increases when traffic load increases; the throughput of the two satellites are slightly unbalanced under light load, but almost equal for heavy load; and the throughputs decrease under heavy load due to the increase in blocking probability. The study also reveals that the throughput results are different under the two models.

### 2.3.2 Other Performance Issues for Low Earth Orbit Satellites

Another study that focuses on analyzing the performance of LEO satellite networks examines the Walker delta pattern configurations (sigma and omega patterns). The performance of the networks has been compared using stochastic Petri net (SPN) modeling schemes and Little's theorem [50]. The main parameters studied include different traffic patterns and mean message delay. Various topologies of low earth orbit satellite networks have been studied [51]. The Walker delta patterns of the satellite constellation, including sigma and omega, are considered superior to the star pattern. The structural properties, such as diameter, mean message traversal, traffic density, maximum message injection rate and message routing algorithm are investigated under three different traffic patterns: uniform message distribution, the local message distribution and the decay probability message distribution.

Using LEO satellites to link cellular networks is an extremely challenging field. The problem of link capacity evaluation in the personal satellite communication system has been discussed in [7]. Personal satellite communication systems should ultimately provide seamless personal communications by joining the cellular networks through Low Earth Orbit (LEO)/ Intermediate Circular Orbit (ICO) satellite systems. For satellites with an orbit height up to 36000 km, determining the number of required full duplex telephone channels between a satellite and its users is very important because it impacts the power and cost of the satellite. By examining the satellite constellations and the traffic distribution on the earth, and performing some simulations, the number of channels can be estimated. Both Low Earth Orbit (LEO) and Intermediate Circular Orbit (ICO) constellations have been examined in terms of number of satellites, number of orbits, phasing factor, inclination, orbit height and minimum guaranteed elevation. Worst case traffic is considered, and the handover criteria to the satellite providing the currently best elevation angle (while minimizing the user handover rate) is adopted. To evaluate the required capacity of the mobile user link, the worst case must be considered, since the required capacity varies with time. The required capacity is analyzed, and a network simulation model is used to estimate the number of channels required with a blocking of at most 5% during peak time. The required number of channels per satellite is much higher for the ICO system than for the LEO system.

## 2.4 Demand Assigned Multiple Access (DAMA)

The Demand Assigned Multiple Access (DAMA) technique allows multiple users to seize the capacity they need from the system, use it, and return it when finished. Due to the merits of this approach, the DAMA multiple access scheme has been extensively researched and applied, and this section examines a few of the results.

The combined Free/Demand Assignment Multiple Access (CFDAMA) Protocol is a class of multiple access schemes that dynamically assign time slots to the user by reservation. This is achieved by using a fixed assigned slot (CFDAMA-FA) or a random access slot (CFDAMA-RA) or by piggybacking a request on a data packet (CFDAMA-PB). In [37], the performance of the CFDAMA-PB scheme is analyzed and simulated to compare it to other multiple access schemes. It is shown that the CFDAMA-PB has a shorter average delay than both the TDMA and TDMA reservation schemes under various loads, and has a lower delay than CRRMA when the load is heavy. The delay variance of CFDAMA-PB is also insensitive to the channel utilization, unlike TDMA and CRRMA.

Another study that employs DAMA proposes a moveable boundary accessing technique for an integrated services multibeam satellite. The integration of video, voice, file data and interactive data is analyzed, using both random access and demand assigned access along with a moving boundary policy. The user population can be substantially increased with the use of the moving boundary policy, and minimal overhead is incurred to accommodate them in the uplink spotbeam. In addition, the scheme substantially improves the performance for interactive data connectionless users which take advantage of the unused connection frame capacity. Analysis reveals that performance of the video services was not strongly dependent on the voice and file users, while they were highly dependent on video traffic. However, it was shown that with proper frame design, a single 150 Mbps spotbeam may accommodate a large number of users [5, 4].

A well-known project that employs the DAMA technique is the Advanced Communication Technology Satellite (ACTS) project. The ACTS performs signal regeneration and switching on-board, hopping and scanning spot beams, Ka-band operation and demand assigned multiple access (DAMA). A test plan has been developed to obtain the approximate value of several of the performance parameters of ACTS for the full experiment of defining the end-to-end digital communication service from the user's perspective. Some of the most important parameters are: access time, access denial probability, block transfer time and block error probability. The test plan consists of three periods and is based on two standards, X3.102 and X3.141 [52].

The ACTS system is also used in [12]. The main operational concepts of a demonstration of a system for emulating satellite cross link communications for command and control operations using the NASA ACTS satellite and VSATs have been explored [12]. The system operates in six modes: roll call status request, roll call response, bandwidth-on-demand, connectivity while tumbling, packet network interfacing and operations in jamming, with different operations for each of them. The major implementation elements for the emulation have also been examined in [12].

## 2.5 Other Multiple Access Schemes

Several other multiple access alternatives have been proposed for satellite networks, and this section is devoted to highlighting the major features of a few of them.

### 2.5.1 Frequency Comb Multiple Access (FCMA)

Frequency Comb Multiple Access (FCMA) is a multiple access method which uses signatures composed of discrete frequency elements to $M-ary$ modulate the data and permit channel access. It is suitable for VSAT networks where only a small number of users are active out of a large number of potential users. On the other hand, VFHMA is a frequency hop satellite multiple access scheme. The performance of FCMA has been studied and compared to VFHMA, as well as pure ALOHA and slotted-ALOHA schemes using an analytical model in [46]. Assuming a binomial distribution as the arrival model, both the throughput (defined as the expected number of correctly received packets in a given time period), and the normalized throughput (defined as the expected number of correctly received packets per Hertz as a function of the packet error rate in a given time period) are derived for the FCMA and VFHMA schemes. The results show that the FCMA scheme is more efficient than the VFHMA scheme, while both of them have better performance than the pure ALOHA and slotted-ALOHA approaches.

### 2.5.2 Announced Retransmission Random Access (ARRA)

Several variations of the Announced Retransmission Random Access (ARRA) scheme for satellite channels have been proposed, where stations announce their retransmission intent by transmitting control information in a mini-slot. Adding this small amount of control information (2% at most) in the data frames significantly increases the capacities, because conflicts can be completely avoided, without compromising the simplicity of random access protocols. The technique is particularly suitable for channels with large propagation delays, such as satellite channels. It is also much simpler than pure ARRA, and does not need expensive circuitry [18].

### 2.5.3 Tone Sense Multiaccess with Partial Collision Detection (TSMA/PCD)

The Tone Sense Multiaccess with Partial Collision Detection (TSMA/PCD) protocol has been proposed for a packet satellite system serving a zone with a dense population of earth stations. A narrowband ground radio channel is used to broadcast busy tones while transmitting packets, so earth stations can avoid packet collisions by sensing the absence of busy tones before transmitting packets. Partial collision detection can also be achieved. Hence, the TSMA/PCD protocol can be used on the uplink, while the downlink can use TDM with statistical multiplexing provided on board. Several variations on TSMA/PCD can be employed: a non-persistent, a 1-persistent or a slot-by-slot announcement TSMA/PCD [36].

## 3 LAN Interconnection

Several systems have been built to investigate the feasibility of the interconnection of local area networks (LANs) by satellites. Issues that arise in such systems include mapping and interfacing of various protocols and formats, suitable routing schemes and flow control mechanisms. The remainder of this section is devoted to highlighting the main aspects of a few of the LAN interconnection satellite systems.

The R2074 project attempts to demonstrate the applicability of satellites in broadband island interconnection across Europe. The island types analyzed included native Asynchronous Transfer Mode (ATM) systems, Distributed Queue Dual Bus (DQDB), Fiber Distributed Data Interface (FDDI), and IEEE 802.3. The project investigates the interconnection of these terrestrial LANs, by analyzing the performance of a variety of client/server architectures using Novell NetWare and Network File System protocols, linked via LAN/ATM/satellite interface units. This will provide a transparent service to a number of ATM islands spanning a wide area, accommodating a variety of traffic demands. The ATM cell size has been observed to be well-suited for satellite transmission, and segmentation of data packets. Forward error correction was employed, and Time Division Multiple Access (TDMA) is chosen as the multiple access scheme, in addition to demand assignment. Traffic shaping was not employed in the simulations, and cells were generated at the rate of 155.5 Mbps, with a worst case traffic of a burst of back-to-back frames. Results showed that efficiency was gained by selecting AAL5 over AAL3/4, but frame delays and response times were not affected [22].

For the interface sub-system, the mapping of CSMA/CD, FDDI and DQDB LAN/MAN frames onto ATM formats is achieved by defining one dedicated LAN/MAN Access Module (LAM) for each type to convert the respective frames to a generic format and then using a Generic LAN/MAN to ATM converter (GLAC) to fragment the generic frames to ATM cells. The GLAC is responsible for implementing the ATM adaptation layer protocol of fragmentation and re-assembly. In addition, it also solves the routing problem by maintaining a database of Virtual Path Identifier (VPI)-LAN/MAN address associations [35].

The same project has been investigated to explore hierarchically grouped routing, and ARQ flow control in [24]. The expected user performance is studied through analysis and simulation using three workstation protocols: stop-and-wait, a window protocol and a high-level blast protocol. The major parameters examined are frame transit time, interactive response time, network throughput and data link utilization.

The architecture and performance of another pilot satellite network is discussed in [3]. The network provides LAN-to-LAN (as an example of high speed data communication) and videoconference (as an example of real-time) applications, and features a meshed topology. A slotted-ALOHA control channel allows the master earth terminal to dynamically assign, on demand (DAMA) or on a permanent mode (PAMA), the satellite capacity to the user earth terminals. The master terminal broadcasts messages to all the user terminals through a TDM channel. On the other direction, the link is supported by one or more burst slotted-ALOHA channels. The traffic subnetwork is based on continuous channels accessing the satellite in SCPC (Single Channel Per Carrier) mode. Experimentations of videoconference applications at a bit rate of $N \times 64$ Kbps ($N = 1, 2, 6$) and LAN-to-LAN interconnection applications between Ethernets at a bit rate of 2-8 Mbps are being performed [3].

Another system for local area network interconnection is the CODE (Co-Operative Data Experiment) system, employing VSATs (Very Small Aperture satellite Terminals). The network connects several cities throughout Europe through routers that connect individual remote LANs to a backbone network provided by the VSATs. Using routers has proved to offer better performance than the traditional way of joining the sites by bridges into a single LAN [19].

For file transfer, packet voice, and some other applications, the performance of TCP/IP in CODE was quite good, but a low throughput has been observed for some common protocols, such as RPC, NFS, and X.11. This is because IP is not suited for links with a large bandwidth delay product, and particularly links with a significant bit error rate (especially during a rain fade). In addition, the IP suite performs most of its functions within the hosts, rather than within the routers, which is not suitable for the VSAT LAN interconnection system. To increase performance, more sophistication must be added to the VSAT router (e.g. using a connection-oriented data link layer with segmentation). Modification of the host protocols (transport and above) might also improve host performance [19].

Hence, the TCP/IP protocol suite operates without modifications over the VSAT link, but applications that rely on a short round trip time achieve minimal performance. In these cases, replacing such applications by ones more suited to VSAT or providing protocol conversion within the VSAT router might be the solution [19].

The performance of TCP/IP for LAN interconnection has also been examined in [9]. The characteristics of delay events are studied for the communication of workstation pairs, using TCP/IP in a full network consisting of FDDI clusters interconnected by satellite links. The analysis results reveal that the window size in the protocol is a crucial parameter for the system performance. A small window size restricts the system and causes a bottleneck for the transmission. It is also shown that a large number of links in a bridge results in long queues, and hence long delays.

# 4 Automatic Repeat Request (ARQ) Protocols

When packets are discarded due to noise or due to congestion, the discarded packets need to be retransmitted. Automatic Repeat Request (ARQ) mechanisms are essential to control packet retransmission, regardless of its cause.

The impact of the long propagation delay of satellite links on the ARQ mechanism has been extensively examined. This section explores a variety of techniques that have been proposed to mitigate that impact. Performance analysis studies have been conducted to verify the proposed techniques, and their results are also summarized here.

## 4.1 Go-Back-N and Selective Reject ARQ

The performance of automatic repeat request (ARQ) protocols in connected-oriented transmission, where each message is divided into several packets, has been examined in [56]. Because connection-oriented systems are considered, messages are assumed to be sent on a First-Come-First-Served basis. For both Go-Back-N and Selective Reject (assuming an infinite size buffer at the receiver), the probability generating functions of message waiting time and queue length are derived. The analysis presented is different from previous work because it handles the cases when messages arriving at the transmitter are addressed to different receivers [56].

A somewhat different Go-Back-N Automatic Repeat Request protocol has been presented and tested in [17]. The protocol requires equal size buffer length in the transmitter and receiver, and uses multicopy retransmissions in its recovery strategy. It avoids buffer overflows, and reduces the effect of the round-trip delay, thus making it suitable for high speed satellite environments, where the channel is good most of the time, and noisy only occasionally. A semi-Markov process model is employed to analyze the protocol, and obtains the throughput as a function of the erroneous messages that may be processed and reordered during the recovery procedure. The analysis shows that the strategy exhibits a high performance, and the throughput efficiency remains in a usable range even for very high error rate conditions. There is a tradeoff to be considered in the implementations where the receiving processing and reordering time is a function of the number of erroneous messages the receiver can process during a recovery procedure.

Another Go-Back-N scheme for error recovery in high speed satellite communications has been especially adapted for point-to-multipoint transmission in [16]. The receivers store the errorless messages received during the recovery procedure. The proposed strategy only uses accumulative acknowledgments and performs very well under a wide variety of data and bit error rates and round trip delays. The accumulative feature not only reduces the number of acknowledgment messages, but also reduces the number of processed control variables for each transmitted information message. The transmitter maintains a state variable for each receiver indicating the number of the last positively acknowledged message. The protocol is analyzed using a discrete-time semi-Markov process, and the throughput efficiency and mean waiting time are measured [16].

## 4.2 Other Retransmission Techniques

Retransmission in broadcast communications has been examined in [29]. A NAK-based broadcast protocol for satellite communications has been proposed, where the link structure and link establishment procedure aim at realizing various broadcast communications without high complexity. Retransmission functions are separated into two layers. The lower sublayer can realize flexible point-to-multipoint communications, while the upper sublayer guarantees error recovery and reliable data transfer.

As previously mentioned, the long propagation delay in satellite networks can adversely influence the retransmission techniques. The tandem type go-back-N scheme has been proposed for satellite communications to reduce the long round-trip delay using the on-board processing of satellites [25]. The throughput performance of the scheme and its application to a broadcast communication system have been analyzed. The study also examines the relationship of the number of receiving stations to the link bit error rate and the round-trip propagation delay. The throughput of the point-to-point tandem Go-Back-N is established to be twice that of the standard scheme. The analytical results are applied to the point-to-multipoint tandem-type Go-Back-N scheme, and the number of receiving stations is shown to be about twice of that for the standard scheme, because of the effective reduction of the round-trip delay.

Another study of the effect of propagation delay on retransmission schemes focuses on TCP. The issue of extending TCP retransmission techniques for use in bandwidth-delay dominated networks, such as satellite networks, has been studied in [28]. Owing to the large propagation delay and high bandwidth, the original algorithms of TCP need to be modified to improve the network performance. It is argued that larger window size is necessary, and data loss during transmission needs special investigation since the cause for the loss, whether it is congestion or noise, implies different network conditions. Three improvements of the ARQ error control and congestion control mechanism are proposed to make the retransmission faster and more efficient. First, NAK is preferred over timeout to speed up the scheme. Second, selective repeat is used to get more efficiency. Third, a strategy named stutter XOR selective repeat is presented, in which the sender does not retransmit NAKed frames immediately upon reception of a NAK, but waits for a certain number of acknowledgments and then decides on retransmitting negatively acknowledged frames. Several NAKed frames are combined together by XORing to minimize the number of retransmissions and to increase throughput.

The idea of combining the frames that need retransmission using XOR has also been proposed in [41]. It is argued that the most efficient ARQ scheme is selective repeat since only erroneous frames are retransmitted, but overflow can be a problem. The proposed scheme is especially designed for the point-to-multipoint scenario, where one retransmission corrects several erroneously transmitted frames at once. It is called the "XOR-protocol" and operates similar to selective repeat. As before, with XOR, the sender does not retransmit NAKed blocks immediately upon the reception of a NAK, but it waits for a certain number of acknowledgments and then decides on retransmitting negatively acknowledged blocks. The idea is to combine several NAKed blocks by XORing them to minimize the number of retransmissions and to increase throughput. The receiver obtains the expected block by applying the XOR operation again to the XOR-block with all the participating and correctly received blocks. The performance gain of this algorithm depends on the number of receivers that take part in the communication. The more receivers are involved, the more the gain will be [41].

# 5 Routing

Due to the flexibility of reconfiguring satellite networks, routing mechanisms that take advantage of this capability have been investigated.

A mesh-connected, circuit-switched satellite communication network is shown to perform better by adaptively reconfiguring the network using simulated annealing to suit the current traffic conditions. Proper allocation of link capacities and placing an optimal reservation scheme of the network are also crucial here, and dynamic routing is employed. Simulation results reveal the improvement in performance when the stochastic method of simulated annealing was used to find a suitable map for the observed traffic conditions. Although dynamic routing improved performance, the simulated annealing method might be slow [1].

# 6 Asynchronous Transfer Mode via Satellites

Employing satellites in a Broadband Integrated Services Digital Network (B-ISDN) imposes severe constraints on the data link and network layer protocols that can be used. The main requirements of using satellites for B-ISDN include: efficient utilization of resources, high transmission rate support, multicast support and transparency to the user. To meet all these requirements, the application of the multi-beam satellite has been proposed in [30]. Some technical issues need to be addressed in this case, such as: multiple access methods, the configuration of the communication system, the channel control schemes, error recovery, congestion control, synchronization, Operation, Administration and Maintenance (OAM), and the interface to the terrestrial network [30]. Some of these issues have already been discussed in the previous sections, and a few others are investigated next.

## 6.1 Error Characteristics

The impact of error characteristics and propagation delay on the operation of B-ISDN via satellites has been investigated in [13], and some solutions are presented to remove adverse effects and provide high quality service. For example, Asynchronous Transfer Mode (ATM) transmission considerations are based upon the assumption that bit errors are randomly distributed. In satellite systems, this assumption will no longer be valid since it is necessary to include channel coding in satellite channels to reduce the receive earth-station size and minimize cost. Channel coding will lead to the occurrence of transmission bit errors in bursts.

## 6.2 Error Correction

Problems regarding reliable data transmission also need to be considered. Error correction can be performed by existing retransmission protocols, but encapsulating this into ATM is limited by scalability and error robustness. Using the service specific connection oriented protocol might be a good option [13].

## 6.3 Congestion Control

Another important issue that is highly complicated by the long propagation delay of the user-satellite channel is the issue of developing congestion control techniques for satellite networks. Congestion control mechanisms are essential in maintaining a specified Quality of Service (QoS) in ATM networks. In satellite networks, due to the large propagation delays, the latency of the feedback mechanisms will be increased. Unless a robust feedback mechanism is designed, the mechanism may become ineffective at a certain point [13]. The next section is devoted to examining the congestion control techniques for satellite networks in more depth.

## 6.4 Multiple Access to ATM Conversion

An additional problem that needs to be addressed for the effective application of B-ISDN systems is the conversion between the ATM systems in the terrestrial B-ISDN and the satellite multiple access scheme. Both [30] and [38] argue that the SS-TDMA multiple access scheme is the most suitable scheme for multiple access in B-ISDN, since the efficiency in the use of the satellite channel is high. They expect that the TDMA system will increase the throughput of satellite channels, as well as accommodate services with various transmission speeds, which renders it the best multiple access method for this system. ATM/TDMA conversion is not a trivial problem, because it increases the Cell Delay Variation (CDV). This problem occurs because cell transmission is only allowed at the preassigned TDMA bursts on the TDMA system. CDV causes buffer over/underflow at exchange/transmission nodes and user terminals. Visual and audio services are especially sensitive to it. Hence a conversion protocol that compensates for that is needed, and [38] proposes such a protocol, namely the cell number counting protocol (CNCP).

Traditional CDV compensation protocols that have been employed on the terrestrial B-ISDN are the leaky bucket method and the timestamp method. However, the leaky bucket method cannot adequately control the absolute CDV value. In addition, the timestamp method, although it reproduces cell intervals completely, causes transmission capacity loss (because the cell arrival interval information consumes about 4% of the total transmission rate), and is very sensitive to bit errors, since it is difficult to recover timing information if there are bit errors in the timestamp information. These problems cannot be ignored in satellite systems [38].

In the cell number counting protocol (CNCP), the input cell stream from the terrestrial network is divided into periods of control time $T_c$ and the average transmission speed in each $T_c$ is represented by the number of cells arriving within $T_c$. At the transmitting earth station, $N$ cells are counted during each $T_c$, and $N$ is sent with the TDMA bursts (thus additional information is negligible). The cell intervals in the receiving station are estimated to be $T_c/N$. Thus the CDV distribution length does not exceed $2 \times T_c$. CDV can be suppressed by setting $T_c$ at an adequate value. The MMPP arrival model is used to accurately represent burst traffic with large fluctuations in transmission speed. Performance analysis of the results reveals that for low speed traffic (less than 3 Mbps) such as audio and visual traffic, the CDV distribution length for $T_c = 1$ ms is less than 1 ms. In addition, the optimization of the $T_c$ parameter value leads to a significant improvement in the CDV compensation performance [38].

## 6.5 SONET/SDH with Satellites

Some researchers argue in favour of using the Synchronous Optical Network (SONET)/Synchronous Digital Hierarchy (SDH) in satellite systems. SDH is one of the most popular means for accomplishing physical layer functions in B-ISDN. The Synchronous Digital Hierarchy (SDH) is suitable for satellite systems, and especially for multi-destination operation. The Plesiochronous Digital Hierarchy (PDH) has several known problems, the most important of which is the problem that multiplexed tributaries are running at slightly different clock rates. The SDH avoids this problem by using pointers to identify the wanted data. For satellite paths, some adaptation is needed before exploiting SDH. First, sub-STM-1 rates are needed for satellite links, and this involves the use of conversion protocols. Second, multi-destination management can take advantage of the simplified SDH multiplexing, but section and path management need to be adapted. Delay and jitter must also be accounted for, but overhead involved in satellite circuits can be reduced with SDH. Preliminary analysis of the system reveals that multi-destination SDH has the promise of being cost-competitive with TDMA, while providing some advantages in terms of utilization and transponders [48].

## 6.6 On-Board Processing

On-board processing satellite packet switches offer improved connectivity, processing gain from demodulation and remodulation, coding gain, and optimized link designs (figure 3) [27].

ISDN connection through an on-board processing satellite system has been explored in [2]. Their system supports both circuit switching and packet switching capacities using multi-frequency TDMA multiple access scheme. The signaling is based on CCITT DSS1, while Segmented scheduled access is used to access the D-channel. For the packet-switching system, DAMA is used to access media, while leaving error control to higher layers in the data network itself. The throughput versus traffic load is studied by simulation with stable results. The next section examines congestion control mechanisms in B-ISDN on-board processing satellites in more detail.

# 7 Congestion Control Mechanisms

Congestion control mechanisms are severely affected by the long propagation delay of the user-satellite link. This problem is more critical in integrated services networks, where bandwidth is allocated on demand, and expected traffic is bursty. Extensive study is still needed before an effective congestion control scheme can be applied in satellite networks. This section attempts to examine the research done on congestion control schemes in satellite networks, focusing on integrated services networks.

In circuit switched satellite networks, congestion is controlled by limiting the access to the network. The problems with this are the inefficient utilization of spectrum and switch capacity, and the long circuit setup and tear down time. On-board packet switching does not suffer from these problems [27].

A satellite packet-switched network must be designed to operate with its own protocols. Hence, satellite virtual packets must be used. Terrestrial-to-satellite protocol converters must be flexible in design [27].

In meshed VSAT networks, transmission is bit-synchronous TDMA up and TDM down. The switch performs both space and time switching. The earth terminal translates incoming data into fixed length packets (2048 bits) to simplify the on-board processing. Each packet is divided into one header and 16 information sub-packets. Forward error correction is employed. Contention is avoided by designing the switch such that its bandwidth exceeds the worst case aggregate up-link data bandwidth. Congestion is controlled through threshold and priorities [27].

In integrated services networks, it is required to multiplex traffic of different Quality of Service (QoS) requirements on a single backbone [40, 55]. Congestion can arise where the network is unable to meet the required QoS for already established connections and/or for new requests. Although a network can avoid this if a new connection is only accepted if the network capacity can accommodate the peak cell rate of this connection (in addition to the peak cell rates of all other established connections), this renders the utilization low. Satellite links are bandwidth limited compared to optical fiber links, so the need for intelligent congestion control mechanisms is even more urgent [13].

In a network with constant bit rate (CBR), variable bit rate (VBR) video and data (such as Asynchronous Transfer Mode networks), the aim of the traffic management scheme is to maintain the high QoS requirement for CBR and VBR, while transporting as much data as possible. 'Preventive controls' in ATM networks include defining traffic more precisely, traffic shaping, 'cell tagging' (setting Cell Loss Priority or CLP), and many other schemes. These mechanisms are inadequate, and reactive mechanisms need to be employed. For example, selective cell discard allows a congested network element to drop non-compliant cells or those with CLP bit set. This, however, leads to problems since dropped cells will need to be retransmitted, and this might lead to excessive delays [13].

Feedback is thus employed so that sources can reduce their load. The network element needs to determine when it is congested, and then it needs to convey that information to the source. The Explicit Forward Congestion Indication (EFCI) mechanism has been employed by altering the payload type in the cell header. When this cell reaches the destination, it informs a higher protocol layer to notify its peer of the congestion. EFCI is not very effective since no semantics have been defined for it, no current user protocol can make use of the indication, it is unenforceable, but, most importantly, it necessarily incurs at least a one way propagation delay in notifying the source. This is unacceptable for satellite networks [13]. It is also noteworthy that the long propagation delay of the satellite links renders congestion avoidance more appealing, since a fixed rate can be used to drain transient bursts.

Backward Explicit Congestion Notification (BECN) might be a better alternative in the case of satellites. A congested network element would send a notification in the reverse direction of the congested path. This notification can be a resource management cell, or any new type of cell. The source would be directed to reduce its rate, and this reduction can be enforced. BECN can achieve a significantly higher performance than FECN in the case of satellite networks. However, for a robust algorithm, network configurations must be considered where congestion occurs on the destination side of the satellite link. The comparative utility of the FECN and BECN mechanisms will vary, depending on the network configuration and the source traffic characteristics. Satellite ATM networks require a high degree of network efficiency with resulting minimal cell loss, and the algorithms to achieve this efficiency must be relatively delay-insensitive. Traffic management specifications still do not adequately address this problem [13].

Other studies on congestion control in satellite networks have been presented in [27, 14]. In [27], a satellite B-ISDN fast packet switch is implemented, and earth terminals translate incoming ATM data into satellite virtual packets. Each packet maybe upto 4 cells long. A QoS field in the packet header is used to prioritize packets on a packet-loss priority basis, while the payload type field is used for delay-sensitivity. Because the B-ISDN switch has a small number of high rate (155 Mbps) users, and the total throughput is greater than 1-2 Gbps, contention and congestion control are accomplished via an output port reservation scheme. Incoming packets are prioritized for delay sensitivity and loss sensitivity and a round-robin system is implemented to ensure fairness among the input formatters. The queues are monitored to detect congestion.

Congestion in the on-board processing switch occurs when the down-link data bandwidth exceeds the down-link transmission bandwidth. The problem is worse because of the long propagation delay (125 ms one way), and is worse for multicasting. The traffic attributes can be divided into 3 layers: the call layer, the burst layer and the packet layer. For each layer, there is a corresponding congestion control strategy [27].

In the call layer, Connection Admission Control (CAC) should be used, by examining the load, service requirements and traffic characteristics. Since the time scale is larger than the propagation delay, link-by-link closed loop negotiation can be implemented. In the packet layer, bandwidth enforcement preventive controls must be used, since the time scale is small. Examples of this include leaky bucket algorithms [27].

The previously described mechanisms do not react to the changing network load, so it is beneficial to add a closed loop reactive control in the burst layer. An example technique is that *the satellite broadcast the buffer status to indicate the onset of congestion and the ground terminals react accordingly.* The status information can be transmitted by a low-rate broadcasting beam or piggyback on data packets. The marking scheme (such as setting the CLP bit) for buffer management is best because it tolerates uncertainty and propagation delay. A combination of the three levels of congestion control is usually the most effective alternative [27].

The scheme described above, where the satellite broadcasts the buffer status to indicate the onset of congestion, is proposed and analyzed in [14]. It is argued that packet-switched satellite networks need to either employ a contention protocol, which is very complex (for the configuration in figure 1), or use a master ground station, which doubles the delay and reduces the bandwidth by half (recall figure 2). Since neither technique is acceptable, *on-board processing* might be the best solution, where the satellite essentially performs all the functions in the data link and network layers (figure 3). A fast data link layer is needed, such as that employing Asynchronous Transfer Mode (ATM) [14].

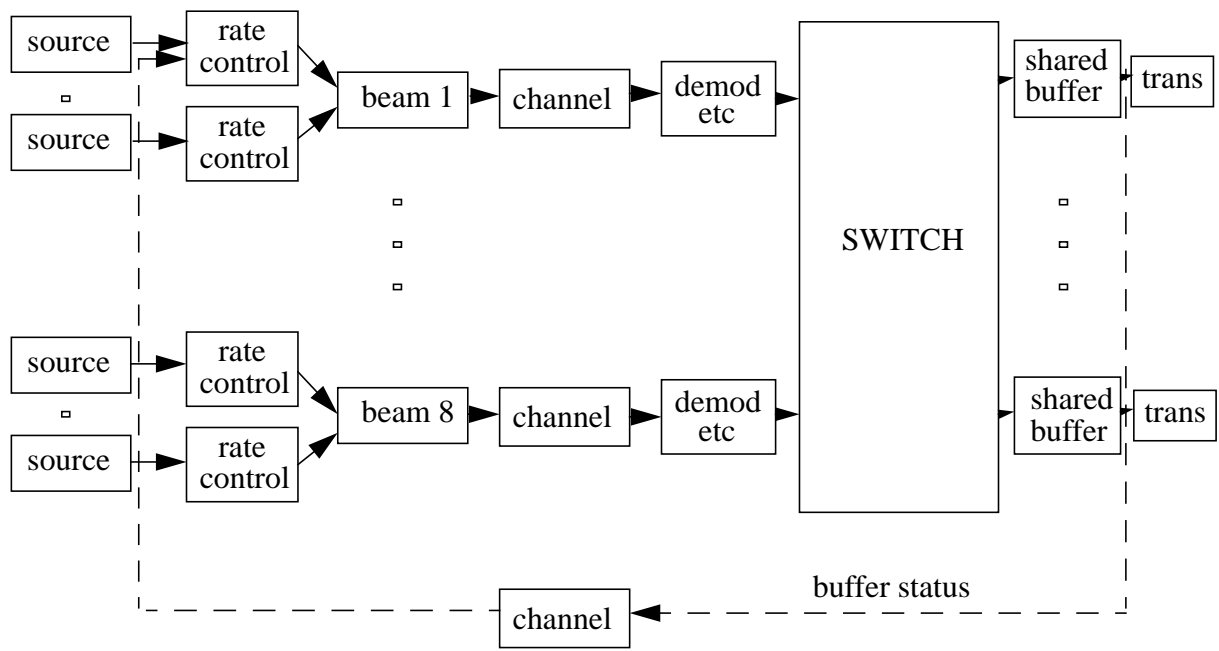

Figure 4: Block Diagram of a Congestion Control Scheme

As previously mentioned, there are two types of controls that can be employed to control congestion: preventive controls and reactive controls. Recall that the long propagation delay in satellite based packet switching networks can severely impact reactive congestion control schemes. The proposed scheme uses a global feedback signal to regulate the packet arrival rate of ground stations. The satellite continuously broadcasts the state of its output buffer. When a ground station detects congestion in the satellite switch, it either reduces its arrival rate by discarding packets, or starts tagging excessive packets as low priority. These low-priority packets will be discarded on-board if congestion actually occurs [14]. A block diagram of the scheme is illustrated in figure 4 (adapted from [14]).

Queueing analysis is used to investigate the impact of propagation delay on the discarding scheme and on the tagging scheme, while simulation results are examined to study the effect of the averaging period on the buffer occupancy, as well as the impact of the reduction function and of tagging [14]. The long propagation delay makes the closed-loop congestion control schemes less responsive, and this can be demonstrated by both analytical and sim-ulation results. The broadcasted information can be used to extract statistical information and perform congestion control, possibly in the burst layer. The discarding scheme needs a lot of ground station discarding to achieve low on-board packet loss probability. Carefully selecting the status information (averaging the buffer occupancy over a period of time) and reduction function improves the performance, but the discarding is still significant. The tagging scheme can be a better alternative since it tolerates more uncertainty caused by the long propagation delay. It can also protect high priority packets from loss, and better utilize the downlink bandwidth. This yields it the most promising scheme. Reactive congestion control can also be incorporated to regulate the incoming traffic for satellite packet switching networks [14].

Testing a congestion control mechanism is quite complicated, because it involves simulating the entire network including the traffic sources and their arrival patterns and channel delay. Developing an exact mathematical model and solving it is almost impossible, and several assumptions need to be made that tend to oversimplify the system. Software simulations are easier to build, but certain aspects of the system might take too long to simulate. Hardware emulation is difficult, but several advances have been made in that respect [27].

Further study is still needed to adequately address the problem of congestion control in satellite networks.

# 8  Summary

Satellite networks suffer from a number of problems that impose severe constraints on the schemes that can be employed to accomplish networking functions. The impact of the long propagation delay in satellite networks has been emphasized when examining various data link and network layer protocols, such as media access control techniques, retransmission protocols and congestion control schemes. Many studies argue that the SS-TDMA multiple access scheme is the most suitable for future broadband integrated services networks, and conversion protocols are needed to convert TDMA to the terrestrial protocol used. Several issues pertaining to media access control techniques have been discussed. In addition, several retransmission protocols have been especially modified to compensate for the long propagation delay of the user-satellite link, and some of them have been highlighted in this survey.

Congestion control techniques for packet switched satellite networks have been carefully examined. We believe that having on-board processing and switching in the satellite usually improves performance, because on-board switches might serve to reduce the effect of the long propagation delay on reactive congestion control schemes. Employing link-by-link closed loop techniques, sometimes referred to as Virtual Source-Virtual Destination might improve the performance of congestion control schemes, because buffer management can be done for each link. The employment of backward explicit congestion notification versus forward explicit congestion notification has also been studied, and a few proposed congestion control schemes have been outlined. The results indicate the need for further study to adequately address the problem of congestion control in satellite networks.

[54] Wen-Bin Yang and Evaggelos Geraniotis. Performance analysis of networks of low-earth-orbit satellites with integrated voice/data traffic. In *MILCOM '93: Communications on the move*, volume 3, pages 978–982, Boston, Mass., 1993. IEEE Communications Society and the Armed Forces, Piscataway, N.J. : IEEE.

[55] Parviz Yegani, Marwan Krunz, and Herman Hughes. Congestion control schemes in prioritized ATM networks. In *Proceedings of the 1994 IEEE International Conference on Communications*, volume 2, pages 1169–1173, IEEE Service Center, Piscataway, NJ, USA, May 1994. IEEE Communications Society, IEEE.

[56] Masakazu Yoshimoto, Tetsuya Takine, Yutaka Takahashi, and Toshiharu Hasegawa. Waiting time and queue length distributions for Go-Back-N and Selective-Repeat ARQ protocols. *IEEE Transactions on Communications*, 41(11):1687–1693, November 1993.31